\documentclass[conference,letterpaper]{IEEEtran}

\usepackage[utf8]{inputenc} 
\usepackage[T1]{fontenc}
\usepackage{url}
\usepackage{ifthen}
\usepackage{cite}
\usepackage[cmex10]{amsmath} 
\usepackage{amsfonts,amssymb}
\usepackage{bm}
\usepackage{bbm}

\newtheorem{dfn}{Definition}
\newtheorem{thm}{Theorem}

\begin{document}
\title{Batch Universal Prediction} 


\author{%
  \IEEEauthorblockN{Marco Bondaschi and Michael Gastpar}
  \IEEEauthorblockA{School of Computer and Communication Sciences\\
                    EPFL\\
                    Switzerland\\
                    Email: \{marco.bondaschi, michael.gastpar\}@epfl.ch}
}


\maketitle


\begin{abstract}
Large language models (LLMs) have recently gained much popularity due to their surprising ability at generating human-like English sentences. LLMs are essentially predictors, estimating the probability of a sequence of words given the past. Therefore, it is natural to evaluate their performance from a universal prediction perspective. In order to do that fairly, we introduce the notion of batch regret as a modification of the classical average regret, and we study its asymptotical value for add-constant predictors, in the case of memoryless sources and first-order Markov sources.
\end{abstract}

\section{Introduction}
Prediction refers to the problem of estimating the next symbols of a sequence given its past, and evaluating the confidence of such an estimate. Such a problem appears in a large number of research areas, such as information theory, statistical decision theory, finance, and machine learning. Some knowledge about the probability distribution that models the sequence to be predicted is clearly helpful. Unfortunately, in many practical applications such knowledge is partial or missing. If this is the case, then one may wish to say something about the future of the sequence when the true model of the source that is producing the symbols is \emph{any} of the models belonging to a certain class. This problem usually goes under the name of \emph{universal prediction} \cite{merhav1}, and it found applications in a wide range of areas, such as compression \cite{ziv1, willems1}, gambling \cite{xie1} and machine learning \cite{fogel1,rosas1}. With the recent rise in popularity of large language models (LLMs), the problem of universal prediction is more timely than ever. In fact, LLMs are essentially predictors: given an $n$-word input sequence, language models can output an estimated probability for the next $\ell$ words in an online fashion. Furthermore, LLMs can be interpreted as \emph{universal} predictors, in the sense that before training, the model does not assume any information about the source generating the input data, and can therefore be used in principle for next-word prediction of data from any distribution, the model improving its prediction accuracy as more and more input data are fed into the network.

Formally, given a finite input alphabet $\mathcal{X}$, for every sequence of symbols $x^i = (x_1,x_2,\dots,x_i)\in\mathcal{X}^i$, a predictor $\hat{p}$ assigns a probability $\hat{p}(y | x^{i})$ for the $(i+1)$-th symbol to be equal to $y\in\mathcal{X}$ given the past $x^{i}$. In universal prediction, a predictor is generally required to perform well if the data is generated according to \emph{any} distribution in a given class $\mathcal{P}$. In order to evaluate the quality of the estimates of a predictor, a \emph{loss function} is used. In language models, this loss function is usually the \emph{logarithmic} (or cross-entropy) loss, which is defined point-wise as 
\begin{equation}
L(\hat{p}, y | x^{i}) = -\log \hat{p}(y|x^{i}).
\end{equation}
In the case $\ell$ sequential symbols have to be predicted, the cumulative loss equals 
\begin{equation}
L(\hat{p}, y^{\ell} | x^i) = -\log \hat{p}(y^{\ell} | x^i), \quad \hat{p}(y^{\ell} | x^i) = \prod_{j=1}^{\ell} \hat{p}(y_j | x^i,y^{j-1}).
\end{equation}
For a given ground-truth distribution $p \in\mathcal{P}$, the \emph{regret} is defined as the difference between the loss of a candidate predictor $\hat{p}$ and that of $p$, i.e., 
\begin{equation}
R(\hat{p},p,y^l |x^i) = L(\hat{p}, y^{\ell} | x^i) - L(p, y^{\ell} | x^i) = \log \frac{p(y^{\ell})}{\hat{p}(y^{\ell} | x^i)}.
\end{equation}
The \emph{average regret} is defined as the expected regret over sequences distributed according to $p$,
\begin{align}
R(\hat{p},p) &= \mathbb{E}_p[R(\hat{p},p,Y^{\ell}|X^i)] \\
	&= \sum_{x^i} p(x^i)\sum_{y^{\ell}}p(y^{\ell})\log \frac{p(y^{\ell})}{\hat{p}(y^{\ell} | x^i)}.
\end{align}
The \emph{maximal average regret} is the maximum average regret over all distributions in $\mathcal{P}$, i.e., $R(\hat{p}) = \max_{p\in\mathcal{P}} R(\hat{p},p)$.

In universal prediction literature, it is customary to consider the regret for the prediction of an entire sequence of $n$ symbols \cite{merhav1, grunwald1}. More precisely, a predictor is defined to output an estimated probability for every $n$-sequence $y^n$, which is denoted by $\hat{p}(y^n)$. The average regret is then $R_n(\hat{p},p) = \sum_{y^n}p(y^n)\log \frac{p(y^n)}{\hat{p}(y^n)}$. This case has been studied extensively, in particular for the memoryless case, where $\mathcal{P}$ is the class of distributions generating i.i.d. symbols. For this case, the asymptotical expression for $R_n(\hat{p},p)$ as $n\to\infty$ has been derived \cite{xie2}. Furthermore, the \emph{add-}$\frac{1}{2}$ predictor, also called Krichevsky-Trofimov predictor, has been shown to be almost asymptotically optimal, in the sense that its asymptotic regret is larger than the optimal one only by a constant independent of $n$. 

However, the rise of LLMs and the importance of viewing them from a universal prediction perspective requires a different paradigm than the one presented above. In fact, LLMs are trained and tested on \emph{batches} of data. In particular, during the training phase a LLM model is fed $n$ batches of data, independent of each other, each of them made of $\ell$ samples. At the end of the training phase, the LLM performance is then measured over a fresh test batch of $\ell$ samples. In order to be able to fairly evaluate LLMs from a universal prediction viewpoint, we introduce a new form of regret, \emph{batch regret}, defined as follows.
\begin{dfn}
Let $\bm{x}^n = (\bm{x}^{(1)},\dots, \bm{x}^{(n)})$ be a training sequence of $n$ batches, each made of $\ell$ samples, $\bm{x}^{(j)} = (x^{(j)}_1,\dots, x^{(j)}_{\ell})$. Let $\hat{p}(y^{\ell} | \bm{x}^n)$ be a predictor that, given $\bm{x}^n$, estimates the probability of a fresh sequence $y^{\ell}$ of $\ell$ samples, generated independently of $\bm{x}^n$. \emph{Batch regret} is then defined as
\begin{equation}
R(\hat{p},p) \triangleq \sum_{\bm{x}^n} p(\bm{x}^{(1)}) \cdots p(\bm{x}^{(n)}) \sum_{y^{\ell}}p(y^{\ell})\log \frac{p(y^{\ell})}{\hat{p}(y^{\ell} | \bm{x}^n)}.
\end{equation}
\end{dfn}

The focus of this paper is to study the asymptotical batch regret for add-constant predictors, in the memoryless and first-order Markov cases. 

\emph{Remark.} We note here that if $n=0$, then we fall back into the classic full-sequence setting described above and described in \cite{merhav1, xie2}. If instead $\ell=1$, we fall into the next-symbol prediction studied by Krichevsky \cite{krichevsky2}. In the full-sequence setting, the Krichevsky-Trofimov predictor is asymptotically the best ``add-constant'' estimator. In the next-symbol setting, instead, the best add-constant estimator requires a slightly larger constant $\beta_0 = 0.50922 \dots$ Our setting can be thought as unifying and generalizing those two perspectives.

\emph{Notation.} The regime of $\ell$ and $n$ that is perhaps most important from a LLM perspective is the one where the batch length $\ell = \ell(n)$ is a function of $n$ such that $\lim_{n\to\infty} \ell(n) = \infty$ and $\lim_{n\to\infty} \frac{\ell(n)}{n} < \infty $. This is the regime of $n$ and $\ell$ considered everywhere in this paper, unless differently specified. The little-o notation $f(n,\ell) = o\left(\frac{1}{n\ell}\right)$ means that $\lim_{n\to\infty} n\ell(n)f(n,\ell(n)) = 0$. All logarithms are considered to be in base $e$.

\subsection{Related work}
The asymptotics of average regret for memoryless sources has been fully characterized in \cite{xie2}. Markov sources from a universal prediction perspective have been studied in \cite{xie1,ryabko1, ryabko2} for the worst-case regret case and recently in \cite{falahatgar1, hao1} for the average regret case. A form of conditional average regret, that matches batch regret in the special case of memoryless sources, was introduced in \cite{liang1}, where it is studied for the location family of distributions. More general forms of regret, defined in terms of R\'enyi divergence, have been studied in \cite{yagli1, bondaschi1} for the memoryless case.

\subsection{Overview}
The remainder of the paper is organized as follows. In Section II we study the binary memoryless case, that is, we consider $\mathcal{P}$ to be the class of distributions generating i.i.d. binary digits. In Section III we study the first-order Markov case, where the next bit is generated according to a distribution that solely depends on the previous bit. In this work we stick to binary alphabet. However, most of the results can be generalized to arbitrary finite alphabets, although proofs become more involved in the general case.

\section{Memoryless sources}
In this section we focus on the following setting. Let $\mathcal{X}=\{0,1\}$ be the binary alphabet, and let $\mathcal{P}$ be the class of memoryless sources, i.e., sources generating i.i.d. binary digits with a given probability,
\begin{equation}
\mathcal{P} = \{p_{\theta}(x^i) = \theta^{n_1} (1-\theta)^{n_0}, \theta \in [0,1], \text{ for any }i\in\mathbb{N}^+\}
\end{equation}
where $n_1$ and $n_0$ are the number of ones and zeros in $x^i$, respectively. In this setting, the average regret is
\begin{multline}
R(\hat{p},\theta) = \sum_{t_1 = 0}^{t} \binom{t}{t_1}\theta^{t_1} (1-\theta)^{t_0} \\
	\sum_{\ell_1=0}^{\ell} \binom{\ell}{{\ell}_1} \theta^{{\ell}_1} (1-\theta)^{{\ell}_0} \log \frac{\theta^{\ell_1} (1-\theta)^{\ell_0}}{\hat{p}(y^{\ell} | x^t)}
\end{multline}
where $t=n\ell$. We note here the performance of two naive predictors. The first predictor is an add-constant predictor that estimates the probability of the new batch $y^{\ell}$ ignoring the past. Such a predictor is defined, for a chosen parameter $\frac{1}{2} \leq \beta \leq 1$, by
\begin{equation} 
\hat{p}(y^{\ell} | x^t) = \prod_{i=1}^{\ell} \hat{p}(y_i | x^t, y^{i-1}),
\end{equation}
where
\begin{equation}
\hat{p}(y_i=1 | x^t, y^{i-1}) = \frac{{\ell}_1^{(i-1)} + \beta}{i-1 + 2\beta}
\end{equation}
and ${\ell}_1^{(i-1)}$ is the number of ones in the sequence $y^{i-1}$.
For such a predictor, the average regret simply becomes the average regret in the full-sequence of length $\ell$ described before. In this case, $\beta = \frac{1}{2}$ achieves the lowest regret, which asymptotically equals $R(\hat{p}_{1/2}) \approx \frac{1}{2}\log \ell + c$. The second naive predictor estimates the probability of a sequence only considering the training sequence $x^t$ and none of the past symbols of $y^{\ell}$. Such a predictor is
\begin{equation}
\label{eq:naive}
\hat{p}_{\beta}(y^{\ell}|x^t) = \left(\frac{t_1 + \beta}{t + 2\beta}\right)^{\ell_1} \left(\frac{t_0 + \beta}{t + 2\beta}\right)^{\ell_0}.
\end{equation}
In this case the regret would be equal to $\ell$ times the Krichevsky next-symbol regret. The best constant would then be $\beta = \beta_0$ and the asymptotic regret would be $R(\hat{p}) \approx \frac{\beta_0 \ell}{t} + o(\frac{1}{t}) = \frac{\beta_0}{n} + o(\frac{1}{n\ell})$. However, both these naive predictors are suboptimal. In this section we study the add-constant predictor that takes into account both $x^t$ and $y^{\ell}$ at the same time. The predictor is defined by
\begin{equation}
\label{eq:pred-def}
\hat{p}_{\beta}(y^{\ell} | x^t) = \prod_{i=1}^{\ell} \hat{p}_{\beta}(y_i | x^t, y^{i-1}),
\end{equation}
where
\begin{equation}
\label{eq:add-beta}
\hat{p}_{\beta}(y_i=1 | x^t, y^{i-1}) = \frac{t_1 + \ell_1^{(i-1)} + \beta}{t + i - 1 + 2\beta}
\end{equation}
for a chosen $\frac{1}{2} \leq \beta \leq 1$. One can think of this predictor as a refined version of the one in \eqref{eq:naive}, where the added constant is updated every time a new symbol from $y^{\ell}$ is revealed. The predictor in \eqref{eq:pred-def} can be rewritten using Gamma functions as
\begin{equation}
\hat{p}_{\beta}(y^{\ell} | x^t) = \frac{\Gamma(t_1+\ell_1+\beta)\Gamma(t_0+\ell_0+\beta)\Gamma(t + 2\beta)}{\Gamma(t_1+\beta)\Gamma(t_0+\beta)\Gamma(t + \ell +2\beta)}
\end{equation}
or again, using properties of the Gamma function, as
\begin{equation}
\hat{p}_{\beta}(y^{\ell} | x^t) = \int_0^1 w_{\beta}(\theta | x^t) p_{\theta}(y^{\ell}) d\theta.
\end{equation}
Hence, the predictor can also be interpreted as a \emph{mixture predictor}, where the prior distribution on the parameter space $w_{\beta}$ depends on the training sequence $x^t$.

For this predictor we have the following results about the batch regret. Theorem 1 concerns the regret in the interior of the simplex, while Theorem 2 concerns the boundary.
\begin{thm}
Let $\delta < \frac{1}{2}$ and $\Theta = [\delta, 1-\delta]$. Let
\begin{equation}
\mathcal{P}_{\delta} = \{p_{\theta}(x^i) = \theta^{n_1} (1-\theta)^{n_0}, \theta \in \Theta, \text{ for any }i\in\mathbb{N}^+\}
\end{equation}
be the class of distributions under consideration. Then,
\begin{align}
\max_{\Theta} R(\hat{p}_{\beta},\theta) &= \frac{1}{2}\log \frac{t+\ell}{t} + o\left(\frac{1}{t}\right) \\
	&= \frac{1}{2}\log \left(1+\frac{1}{n}\right) + o\left(\frac{1}{n\ell}\right)
\end{align}
\end{thm}
\begin{IEEEproof}
The following proof follows and extends the one in \cite{xie2}. We start by rewriting the regret as
\begin{multline}
\label{eq:R2}
R(\hat{p}_{\beta},\theta) = \sum_{z_1=0}^{z} \binom{z}{z_1} \theta^{z_1}(1-\theta)^{z_0} \log\frac{\theta^{z_1}(1-\theta)^{z_0}}{\frac{\Gamma(z_1+\beta)\Gamma(z_0+\beta)}{\Gamma(z+2\beta)}} \\
	-\sum_{t_1=0}^{t} \binom{t}{t_1} \theta^{t_1}(1-\theta)^{t_0} \log\frac{\theta^{t_1}(1-\theta)^{t_0}}{\frac{\Gamma(t_1+\beta)\Gamma(t_0+\beta)}{\Gamma(t+2\beta)}}
\end{multline}
where $z = t+\ell$. We analyze the first term in the equation. The second term can be analyzed similarly. We will extensively use the equality 
\begin{equation}
\Gamma(x) = \sqrt{2\pi} x^{x-\frac{1}{2}} e^{-x} e^s
\end{equation}
for some $s\in [0,\frac{1}{12x}]$. Using this equation in the first term of the right-hand side of \eqref{eq:R2} we get
\begin{align*}
\sum_{z_1=0}^{z} &\binom{z}{z_1} \theta^{z_1}(1-\theta)^{z_0} \log\frac{\theta^{z_1}(1-\theta)^{z_0}}{\frac{\Gamma(z_1+\beta)\Gamma(z_0+\beta)}{\Gamma(z+2\beta)}} \\
	&= \sum_{z_1=0}^{z} \binom{z}{z_1} \theta^{z_1}(1-\theta)^{z_0} \log \Bigg\{\sqrt{\frac{(z_1 + \beta)(z_0+\beta)}{2\pi (z+2\beta)}} \\
	&\qquad\cdot\theta^{z_1}(1-\theta)^{z_0}\left(\frac{z_1+\beta}{z+2\beta}\right)^{-z_1 - \beta} \left(\frac{z_0+\beta}{z+2\beta}\right)^{-z_0 - \beta} \\
	&\hspace{14em}\cdot e^{s(z) - s_1(z_1) - s_2(z_0)}\Bigg\}
\end{align*}
We can split the last expression into three terms that will be analyzed separately:
\begin{multline}
-\sum_{z_1=0}^{z} \binom{z}{z_1} \theta^{z_1}(1-\theta)^{z_0} \left(z_1+\beta-\frac{1}{2}\right)\log (z_1+\beta) \\
	- \sum_{z_0=0}^{z} \binom{z}{z_0} \theta^{z_1}(1-\theta)^{z_0}\left(z_0+\beta-\frac{1}{2}\right)\log (z_0+\beta) \tag{A}
\end{multline}
\begin{equation}
-\frac{1}{2}\log(2\pi) + \left(z+2\beta - \frac{1}{2}\right)\log (z+2\beta) \tag{B}
\end{equation}
\begin{equation}
\sum_{z_1=0}^{z} \binom{z}{z_1} \theta^{z_1}(1-\theta)^{z_0} (s(z) - s_1(z_1) - s_0(z_0)) \tag{C}
\end{equation}
For term (A), consider only the first half (the second follows by symmetry), with the positive sign for simplicity. We can split it into two further terms
\begin{equation}
\sum_{z_1=0}^{z} \binom{z}{z_1} \theta^{z_1}(1-\theta)^{z_0} \left(z_1+\beta-\frac{1}{2}\right)\log \left(z_1+\beta - \frac{1}{2}\right) \tag{A'}
\end{equation}
\begin{multline}
\sum_{z_1=0}^{z} \binom{z}{z_1} \theta^{z_1}(1-\theta)^{z_0} \left(z_1+\beta-\frac{1}{2}\right)\\
	\log \left(1+\frac{1}{2(z_1 + \beta - \frac{1}{2})}\right) \tag{A''}
\end{multline}
We analyze term (A'), which equals $\mathbb{E}[(Z_1 + \beta - \frac{1}{2}) \log(Z_1 + \beta - \frac{1}{2})]$ for $Z_1 \sim \mathrm{Binomial}(z,\theta)$ . Using the bounds (valid for $a \geq 0$ and $b > 0$)
\begin{equation}
a \log a \geq b \log b + (a-b)(1+\log b) + \frac{(a-b)^2}{2b} - \frac{(a-b)^3}{6b^2}
\end{equation}
and
\begin{multline}
a \log a \leq b \log b + (a-b)(1+\log b) + \frac{(a-b)^2}{2b} \\
	- \frac{(a-b)^3}{6b^2} + \frac{(a-b)^4}{3z^3}
\end{multline}
with $a = Z_1 + \beta - \frac{1}{2}$ and $b = z\theta + \beta - \frac{1}{2}$, we have
\begin{align}
\mathbb{E}&\left[\left(Z_1 + \beta - \frac{1}{2}\right) \log\left(Z_1 + \beta - \frac{1}{2}\right)\right] \notag \\
	&\geq \left(z\theta + \beta - \frac{1}{2}\right)\log\left(z\theta + \beta - \frac{1}{2}\right) \notag \\
	&\quad+ \frac{1-\theta}{2} - \frac{1}{z\theta}
\end{align}
and 
\begin{align}
\mathbb{E}&\left[\left(Z_1 + \beta - \frac{1}{2}\right) \log\left(Z_1 + \beta - \frac{1}{2}\right)\right] \notag \\
	&\leq \left(z\theta + \beta - \frac{1}{2}\right)\log\left(z\theta + \beta - \frac{1}{2}\right) \notag \\
	&\quad+ \frac{1-\theta}{2} + \frac{1}{z\theta}
\end{align}
for $z$ large enough such that $z\theta \geq 2$.
For the term (A'') we have
\begin{equation}
\frac{1}{2} - \frac{1}{2z\theta}\leq \mathbb{E}\left[\left(Z_1 + \beta - \frac{1}{2}\right)\log\left(1 + \frac{1}{2(Z_1 + \beta - \frac{1}{2})}\right)\right] \leq \frac{1}{2}
\end{equation}
which follows from the fact that for $a \geq 0$,
\begin{equation}
\frac{1}{2} - \frac{1}{2(a+1)} \leq a\log \left(1+\frac{1}{2a}\right) \leq \frac{1}{2}
\end{equation}
and from the fact that $\mathbb{E}[1/(Z_1+1)] \leq 1/z\theta$. The (straightforward) analysis of terms (B) and (C) is left in the Appendix.
The same analysis worked out for the first term of \eqref{eq:R2} can be carried out for the second term as well, just with $t$ in place of $z$. Putting the bounds together leads to the theorem.
\end{IEEEproof}

\begin{thm}
Let $\theta \in \{0,1\}$. Then
\begin{align}
R(\hat{p}_{\beta},\theta) &= \beta\log \frac{t+\ell}{t} + o\left(\frac{1}{t}\right) \\
	&= \beta\log \left(1+\frac{1}{n}\right) + o\left(\frac{1}{n\ell}\right)
\end{align}
\end{thm}
\begin{IEEEproof}
See the Appendix.
\end{IEEEproof}
Note that Theorem 1 and 2 are not enough to prove the exact asymptotics of the batch regret in the case the class of distributions is parametrized by the entire interval $\Theta = [0,1]$. How the regret behaves in this case is open and left for future work.

\section{First-order Markov sources}
In this section we deal with the class of binary Markov sources of order 1, i.e., the class
\begin{equation}
\label{eq:markov-class}
\mathcal{P} = \left\{p_{\theta}(x^i) = p_1(x_1)\prod_{j=2}^i p(x_j | x_{j-1})\right\}.
\end{equation}
For notation purposes we define $p_1 = p_1(x_1 = 1)$, $p = p(1|0)$, $q = p(0|1)$ and $\theta = (p_1,p,q) \in [0,1]^3$.
While in traditional full-sequence universal prediction or in the Krichevsky next-symbol prediction, the value of the initial distribution $p_1$ is asymptotically unimportant, in the batch prediction setting described in this paper it gets a more prominent role. In fact, the training data is a collection of $n$ fresh batches of $\ell$ bits each, where each batch $\bm{x}^{(i)}$ is generated independently of the others according to a fixed Markov source, i.e.,
\begin{equation}
p_{\theta}(\bm{x}^{(i)}) = p_1(x_1^{(i)}) \prod_{j=2}^{\ell} p(x_j^{(i)} | x_{j-1}^{(i)}).
\end{equation}
Due to this different feature, depending on the regime of $\ell$ and $n$, the starting distribution $p_1$ becomes fundamental. Hence, we must introduce an estimator for the starting distribution $p_1$ as well. For a predictor in the form
\begin{equation}
\hat{p}(y^{\ell} | \bm{x}^n) = \hat{p}_1(y_1 | \bm{x}^n) \hat{p}(y_2^{\ell} | \bm{x}^n, y_1)
\end{equation}
the batch regret equals
\begin{align}
R(\hat{p},\theta) &= \sum_{\bm{x}^n} p_{\theta}(\bm{x}^n) \cdots p_{\theta}(\bm{x}^{(n)}) \sum_{y^{\ell}}p_{\theta}(y^{\ell})\log \frac{p_{\theta}(y^{\ell})}{\hat{p}(y^{\ell} | \bm{x}^n)} \notag\\
	&= \sum_{\bm{x}^n} p_{\theta}(\bm{x}^n) \sum_y p_1(y) \log \frac{p_1(y)}{\hat{p}_1(y|\bm{x}^n)} \notag \\
		&+ \sum_{\bm{x}^n} p_{\theta}(\bm{x}^n) \sum_y p_1(y) \sum_{y_2^{\ell}} p_{\theta}(y_2^{\ell}|y_1) \log \frac{p_{\theta}(y_2^{\ell}|y_1)}{\hat{p}_{\beta}(y_2^{\ell}|\bm{x}^n,y)}
\end{align}
where $p_{\theta}(\bm{x}^n) = \prod_{i=1}^n p_{\theta}(\bm{x}^{(i)})$. The last expression shows that the regret can be seen as the sum of two terms: the first term is the regret for the estimation of the initial distribution, while the second term is the regret for the estimation of the transition probabilities of the Markov source. We now derive upper bounds for the two terms for Markov sources with positive transition probabilities. Both bounds are asymptotically proportional to $\frac{1}{n}$, so that neither of them can be neglected.
\begin{thm}
Let $\mathcal{P}$ be the class of first-order binary Markov sources as in \eqref{eq:markov-class}. The initial distribution regret
\begin{equation}
R_1(\hat{p},\theta) = \sum_{\bm{x}^n} p_{\theta}(\bm{x}^n) \sum_y p_1(y) \log \frac{p_1(y)}{\hat{p}_1(y|\bm{x}^n)}
\end{equation}
is upper bounded by
\begin{equation}
\min_{\hat{p}_1}\max_{\theta} R_1(\hat{p},\theta) \leq \frac{\beta_0}{n} + o\left(\frac{1}{n}\right).
\end{equation}
\end{thm}
\begin{IEEEproof}
Let $\bm{x}_1 = (x_1^{(1)},x_1^{(1)}, \dots, x_1^{(n)})$ be the sequence of bits in the first coordinate of each batch, and consider the predictor
\begin{equation}
\hat{p}_1(y | \bm{x}^n) = \hat{p}_1(y | \bm{x}_1) =\frac{t_1 + \beta_0}{n + 2\beta_0}
\end{equation}
where $t_1 = \sum_{i=1}^n x_1^{(i)}$ is the number of ones in $\bm{x}_1$. Then,
\begin{equation}
R_1(\hat{p},\theta) = \sum_{\bm{x}_1} p_1(\bm{x}_1)  \sum_y p_1(y) \log \frac{p_1(y)}{\hat{p}_1(y|\bm{x}_1)}.
\end{equation}
The last expression is precisely Krichevsky's next-symbol regret for the add-$\beta_0$ predictor. Hence, Krichevsky's bound can be applied, leading to
\begin{equation}
\max_{\theta} R_1(\hat{p},\theta) \leq \frac{\beta_0}{n} + o\left(\frac{1}{n}\right).
\end{equation}
\end{IEEEproof}
Furthermore, if we limit the class of predictors to those than only depend on the first coordinate of each batch, $\bm{x}_1 = (x_1^{(1)},x_1^{(1)}, \dots, x_1^{(n)})$, then the following lower bound on $R_1(\hat{p},\theta)$ follows directly from Krichevsky's lower bound on next-symbol prediction \cite[Theorem 2]{krichevsky2}.
\begin{thm}
Let $\hat{\mathcal{P}}$ be the class of predictors $\hat{p}_1(y | \bm{x}^n) = \hat{p}_1(y | \bm{x}_1)$ that only depend on $\bm{x}_1$. Then,
\begin{equation}
\min_{\hat{p}_1 \in \hat{\mathcal{P}}}\max_{\theta} R_1(\hat{p},\theta) \geq \frac{1}{2n} + o\left(\frac{1}{n}\right).
\end{equation}
\end{thm}

\emph{Remark.} Theorem 3 suggests that one should use an add-$\beta_0$ predictor with the first coordinate of the $n$ batches when predicting the initial distribution of a Markov source. However, one can possibly achieve lower regret with a predictor that also uses the other coordinates of the batches to estimate the initial distribution of the Markov source. In fact, information about $p_1$ also leaks to other coordinates. Note that the second symbol of each batch is distributed according to 
\begin{equation}
\mathrm{Pr}(X_2 = 1) = p_1 (1-p-q) + p.
\end{equation}
By inverting the last expression, one gets
\begin{equation}
p_1 = \frac{\mathrm{Pr}(X_2 = 1) - p}{1 - p -q}
\end{equation}
Therefore, one can combine estimators for $\mathrm{Pr}(X_2 = 1)$, $p$ and $q$ to derive a predictor for $p_1$. Obvious choices would be to estimate $\mathrm{Pr}(X_2 = 1)$ with an add-constant estimator based on the counts in the second coordinate of the $n$ batches $\bm{x}^n$, while add-constant predictors based on the counts of transitions in the entire training set could be used to estimate $p$ and $q$. One can derive similar predictors for $p_1$ from the other coordinates as well, since in general ${\mathrm{Pr}(X_j = 1) = (p_1 - \pi_1)(1-p-q)^{j-1} + \pi_1}$, where $\pi_1 = \frac{p}{p+q}$ is the probability assigned to $1$ by the stationary distribution of the Markov source. It is therefore natural to average those $\ell$ predictors of $p_1$ to achieve lower regret than the one obtained by only considering the first coordinate of the batches.

The following is an upper bound for the regret on the transition probabilities of the Markov source, provided that all transition probabilities are bounded away from zero.
\begin{thm}
Let $0 < \delta < \frac{1}{2}$ and let $\mathcal{P}_{\delta}$ be the class of first-order binary Markov sources as in \eqref{eq:markov-class}, such that $p,q \in [\delta, 1-\delta]$. The transition probability regret
\begin{equation}
R_{\rm T}(\hat{p},\theta) = \sum_{\bm{x}^n} p_{\theta}(\bm{x}^n) \sum_{y^{\ell}} p_{\theta}(y^{\ell}) \log \frac{p_{\theta}(y_2^{\ell} | y_1)}{\hat{p}(y_2^{\ell}|\bm{x}^n,y_1)}
\end{equation}
is upper-bounded by
\begin{equation}
\min_{\hat{p}} \max_{p_{\theta} \in \mathcal{P}_{\delta}} R_{\rm T}(\hat{p},\theta) \leq \frac{1}{2n} + o\left(\frac{1}{n\ell}\right) .
\end{equation}
\end{thm}
\begin{IEEEproof}
Consider the predictor
\begin{equation}
\label{eq:pred-markov-2}
\hat{p}(y_2^{\ell} |y_1) = \prod_{h,k \in \{0,1\}^2} \left(\frac{t_{hk} + \frac{1}{2}}{t_h + 1}\right)^{\ell_{hk}} .
\end{equation}
Let $\pi$ be the stationary distribution of the Markov source. Note that $\mathbb{E}[L_{hk}] = \ell \pi(h) p(k|h) + o(\frac{1}{\ell})$. Furthermore, we have 
\begin{align}
\mathbb{E}[T_{hk}] &= \sum_{i=1}^n \mathbb{E}[T_{hk}^{(i)}] \\
	&= n\ell \pi(h) p(k|h) + o(n/\ell)
\end{align}
Hence, we can write
\begin{align}
R_{\rm T}(\hat{p},\theta) &= \mathbb{E}\left[\sum_{hk} L_{hk}\log \frac{p(k|h)}{\frac{T_{hk} + \frac{1}{2}}{T_h + 1}}\right] \\
	&= \sum_{hk} \mathbb{E}[L_{hk}] \mathbb{E}\left[\log \frac{p(k|h)}{\frac{T_{hk} + \frac{1}{2}}{T_h + 1}}\right] \\
	&= \ell \sum_h \pi(h) \mathbb{E}\left[\sum_k p(k|h) \log \frac{p(k|h)}{\frac{T_{hk} + \frac{1}{2}}{T_h + 1}}\right]+ o\left(\frac{1}{n\ell}\right). \label{eq:RT-bound}
\end{align}
The expectation in the last expression is the \emph{estimation risk} denoted by $\tilde{\epsilon}$ in \cite{hao1}, except for one key difference: here the expectation is over $n$ independent batches of length $\ell$ each, while in \cite{hao1} it is over one batch of length $n \ell $. However, the proof provided in \cite[Section 10]{hao1} also works in the batch case presented here. In fact, the key ingredient in the original proof is the fact that $T_h$ and $T_{hk}$ are highly concentrated around their mean. The same holds in the batch case, since $T_h$ and $T_{hk}$ are sums of independent, highly concentrated random variables: applying \cite[Lemma 19]{hao1} and Hoeffding's inequality to $T_{hk}$ and $T_h$ gives the concentration bound $\mathrm{Pr}(| T_h - n\ell\pi(h)| > t) \lesssim \exp(-t^2 / n\ell)$. Hence, one can prove that
\begin{equation}
\mathbb{E}\left[\sum_k p(k|h) \log \frac{p(k|h)}{\frac{T_{hk} + \frac{1}{2}}{T_h + 1}}\right] \leq \frac{1}{2n\ell} + o\left(\frac{1}{n\ell}\right).
\end{equation}
Putting this into \eqref{eq:RT-bound} leads to the theorem.
\end{IEEEproof}

Similarly as the bound on the regret for the initial distribution, also in this case a lower bound proportional to $\frac{1}{n}$ can be derived if one limits the class of predictors to those in the form \eqref{eq:pred-markov-2}. Using also the latter information can lead to a lower regret. A natural modified version of the add-constant predictor for Markov sources that uses all available information takes the form
\begin{equation}
\label{eq:pred-markov}
\hat{p}_{\beta}(y^{\ell} | \bm{x}^n) = \hat{p}_1(y_1|\bm{x}^n)\prod_{j=2}^{\ell} \hat{p}_{\beta}(y_j| \bm{x}^n, y^{j-1}),
\end{equation}
where 
\begin{equation}
\hat{p}_{\beta}(y_j=k | \bm{x}^n, y^{j-2}, y_{j-1}=h) = \frac{t_{hk} + \ell_{hk}^{(j-1)} + \beta}{t_h + \ell_h^{(j-1)} + 2\beta}
\end{equation}
where for $h,k\in\{0,1\}$ we define
\begin{align}
t_{hk} &= \sum_{i=1}^n t_{hk}^{(i)}\ , \quad t_{hk}^{(i)} = \text{n. of consecutive } hk \text{ in } \bm{x}^{(i)} \\
t_{h} &= \sum_{i=1}^n t_{h}^{(i)}\ , \quad t_{h}^{(i)} = \text{n. of } h \text{ in } \bm{x}^{(i)} \setminus x_{\ell}^{(i)}
\end{align}
and where $\ell_{hk}^{(j-1)}$ is the number of consecutive $hk$ in $y^{j-1}$, while $\ell_h^{(j-1)}$ is the number of  $h$ in $y^{j-2}$.
Furthermore, $\hat{p}_1(y_1|x^t)$ is the predictor for the initial distribution $p_1$. However, we conjecture that one cannot find a predictor with a regret that decays asymptotically faster than $\frac{1}{n}$, even if the multiplying constant of the leading term may improve.

\IEEEtriggeratref{9}


\bibliographystyle{IEEEtran}
\bibliography{prediction}

\newpage
\appendix
\subsection{Analysis of terms (A) and (B) for the proof of Theorem 1}
For term (B), we can use the bound
\begin{equation}
\frac{x}{1+x} \leq \log (1+x) \leq x
\end{equation}
valid for $x > -1$, to easily get
\begin{multline}
\frac{1}{2}\log(2\pi) - \left(z+2\beta - \frac{1}{2}\right)\log (z+2\beta) \\
	\leq \frac{1}{2}\log(2\pi) - z\log z - \left(2\beta - \frac{1}{2}\right)\log z - 2\beta + \frac{\beta}{z}
\end{multline}
and
\begin{multline}
\frac{1}{2}\log(2\pi) - \left(z+2\beta - \frac{1}{2}\right)\log (z+2\beta) \\
	\geq \frac{1}{2}\log(2\pi) - z\log z - \left(2\beta - \frac{1}{2}\right)\log z - 2\beta - \frac{3\beta}{z}.
\end{multline}
Finally, for term (C) we have 
\begin{equation}
\mathbb{E}[s(z) - s_1(Z_1) - s_0(Z_0)] \leq \frac{1}{12z}
\end{equation}
and 
\begin{align}
\mathbb{E}[s(z) - s_1&(Z_1) - s_0(Z_0)] \\
	&\geq -\mathbb{E}\left[\frac{1}{12(Z_1 + \beta)}\right] -\mathbb{E}\left[\frac{1}{12(Z_0 + \beta)}\right] \\
	&\geq -\frac{1}{12z\theta} - \frac{1}{12z(1-\theta)}.
\end{align}
where in the last step we used $\mathbb{E}\left[\frac{1}{Z_1 + 1}\right] \leq \frac{1}{z\theta}$.

\subsection{Proof of Theorem 2}
We prove the case $\theta = 1$. The case $\theta = 0$ follows in the same way due to symmetry. We have
\begin{align}
R(\hat{p},\theta=1) &= \log \frac{\Gamma(t+\ell+2\beta)\Gamma(t+\beta)}{\Gamma(t+\ell+\beta)\Gamma(t+2\beta)} \\
	&= \log\frac{\Gamma(t+\ell+2\beta)}{\Gamma(t+\ell+\beta)} + \log\frac{\Gamma(t+\beta)}{\Gamma(t+2\beta)}.
\end{align}
Using the double inequality
\begin{equation}
 \frac{x}{(x+s)^{1-s}}\leq \frac{\Gamma(x+s)}{\Gamma(x)} \leq x^s
\end{equation}
valid for $x>0$ and $0<s<1$, we get the upper bound
\begin{align}
R(\hat{p},\theta=1) &\leq \beta\log (t+\ell+\beta) + \log \frac{(t+2\beta)^{1-\beta}}{t+\beta} \\
	&\leq \beta\log\frac{t+\ell}{t} + \frac{\beta^2}{t+\ell} + \frac{2\beta(1-\beta)}{t}
\end{align}
and the lower bound
\begin{align}
R(\hat{p},\theta=1) &\geq \log\frac{t+\ell+\beta}{(t+\ell+2\beta)^{1-\beta}} -\beta\log(t+\beta)\\
	&\geq \beta\log\left(\frac{t+\ell}{t}\right) - \frac{\beta}{t+\ell} -\frac{\beta^2}{t}.
\end{align}
These two bounds together prove the theorem.
\end{document}